\documentclass[prb,twocolumn,amsmath,amssymb]{revtex4}

\usepackage{graphicx}
\usepackage{bm}
\usepackage{gensymb}
\usepackage{multirow}
\usepackage{xcolor}
\usepackage{csquotes}
\usepackage{braket}
\usepackage{hyperref} 
\usepackage{soul}
\DeclareGraphicsExtensions{.eps, .jpg,.pdf,.png}

\begin{document}

\title{CeIr$_{3}$Ge$_{7}$: a local moment antiferromagnetic metal with extremely low ordering temperature}
	
\author{Binod K. Rai$^1$}
\author{Jacintha Banda$^2$}
\author{Macy Stavinoha$^3$}
\author{R. Borth$^2$}
\author{D.-J. Jang$^2$}
\author{Katherine A. Benavides$^4$}
\author{D. A. Sokolov$^2$}
\author{Julia Y. Chan$^4$}
\author{M. Nicklas$^2$}
\author{Manuel Brando$^2$}
\author{C.-L. Huang$^1$}
\author{E. Morosan$^1$}
	
\affiliation{$^1$Department of Physics and Astronomy, Rice University, Houston, TX 77005 USA
\\$^2$Max Planck Institute for Chemical Physics of Solids, Dresden, 01187 Germany
\\$^3$ Department of Chemistry, Rice University, Houston, TX 77005 USA
\\$^4$ Department of Chemistry \& Biochemistry, University of Texas at Dallas, Richardson, TX 75080 USA}
\date{\today}

	\begin{abstract}
CeIr$_3$Ge$_7$ is an antiferromagnetic metal with a remarkably low ordering temperature $T_{\rm N}$ = 0.63 K, while most Ce-based magnets order between 2 and 15 K. Thermodynamic and transport properties as a function of magnetic field or pressure do not show signatures of Kondo correlations, interaction competition, or frustration, as had been observed in a few antiferromagnets with comparably low or lower $T_{\rm N}$. The averaged Weiss temperature measured below 10 K is comparable to $T_{\rm N}$ suggesting that the RKKY exchange coupling is very weak in this material. The unusually low $T_{\rm N}$ in CeIr$_3$Ge$_7$ can therefore be attributed to the large Ce-Ce bond length of about 5.7 {\AA}, which is about 1.5 {\AA} larger than in the most Ce-based intermetallic systems. 
	\end{abstract}
	
	\maketitle
	
	
Compounds containing Ce or Yb ions have been studied extensively due to their diverse ground states originating from the competition between several energy scales. The competition between Ruderman-Kittel-Kasuya-Yosida (RKKY) exchange interaction and Kondo coupling give rise to intermediate valence behavior\cite{Currat1989,Mazet2013, BinodAPL, BinodPRB} or Kondo screening\cite{Morosan2006, GegenwartandSi2007, PaglioneandMaple2007,BinodPRB,BinodNM2017,BinodPRX2017} which, in turn, often result in unconventional superconductivity, non-Fermi liquid behavior, and quantum criticality\cite{Loehneysen, Schuberth and Steglich, PaglioneandMaple2007, Custers and Coleman,Steppke and Brando}. If the hybridization between $f$ electrons and conduction electrons is very weak, the ground states of these systems are dictated by RKKY exchange interaction and crystal electric field (CEF) effects, resulting in long-range magnetic order.\cite{Fritsch2011, Huang2012, Ajeesh2017} In the case of the Ce local moment metals without the Kondo effect, the ordering temperatures range from $T_{\rm C}$ = 115 K\cite{Dhar1981,Givord2007} in ferromagnetic CeRh$_3$B$_2$, to $T_{\rm N}$ = 2.75 K \cite{Schoop} in antiferromagnetic (AFM) CeSbTe, while much lower temperatures can be expected for the Yb analogues\cite{Flouquet}. Lower ordering temperatures in both Ce$^{3+}$ and Yb$^{3+}$ compounds could occur from any combination of effects including Kondo, competition between different exchange interactions, strong CEF anisotropy, or large distances between rare-earth ions ($d_{R-R}$) that minimize the RKKY exchange coupling $J_{RKKY}$. Compounds with low ordering temperatures often involve either weaker-than-RKKY exchange, as is the case in insulators, or  multipolar order \cite{Portnichenko}, in which case the resulting order is almost always underlined by heavy fermion (HF) behavior. A remarkably low N{\'e}el temperature $T_{\rm N}$ = 0.18 K has been observed in the intermetallic compound Ce$_4$Pt$_{12}$Sn$_{25}$ \cite{Kurita}. However, the peak around $T_{\rm N}$ in the magnetic specific heat extends well into the paramagnetic state, and this has been attributed to either the onset of Kondo screening or frustration.
	
	\begin{figure}[ht!]
		\includegraphics[width=1\columnwidth]{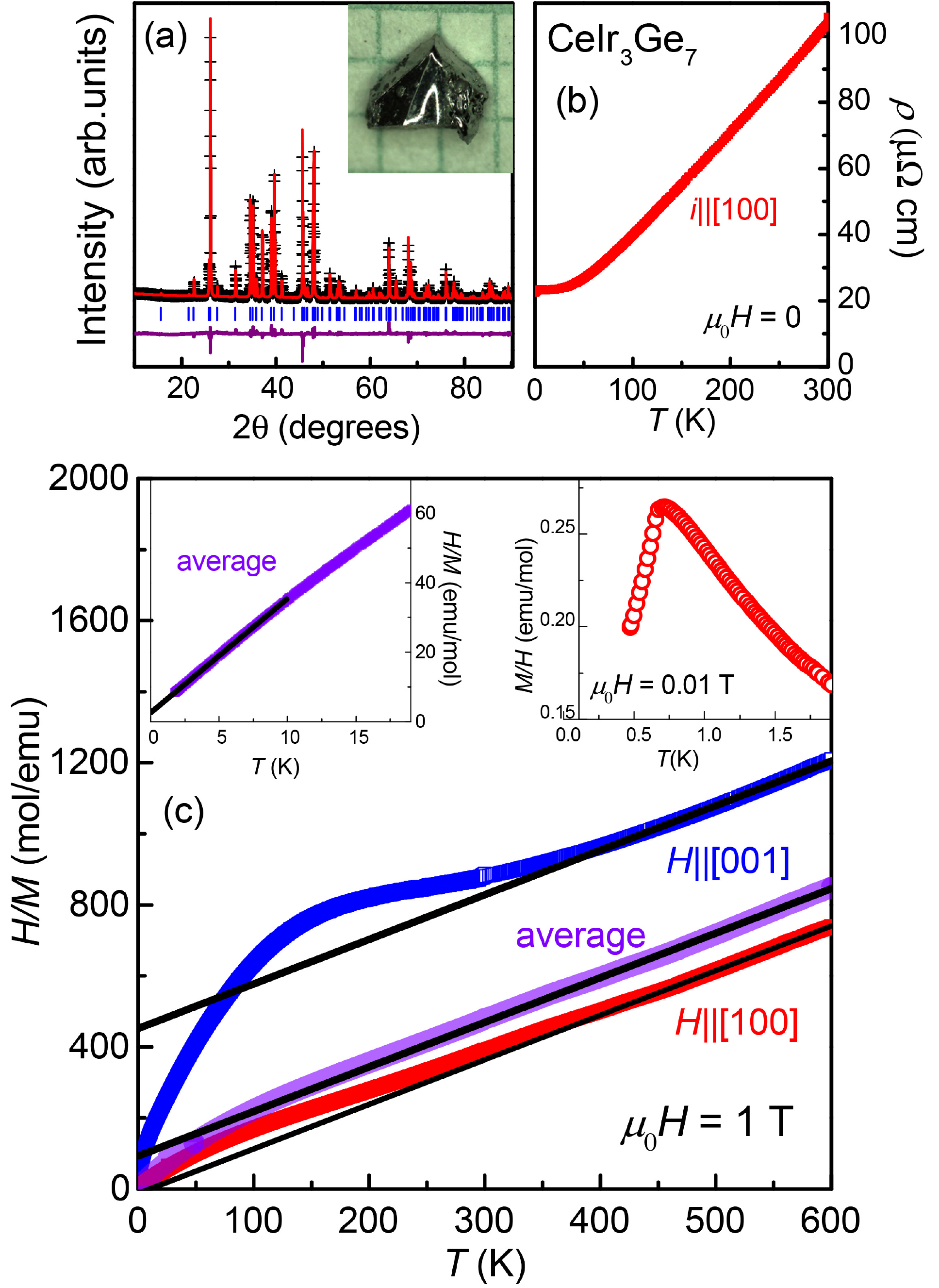}
		\caption{\label{Fig1} (a) Room temperature (symbols) and calculated (red line) powder x-ray diffraction patterns of CeIr$_3$Ge$_7$, together with the expected peak positions (blue vertical lines) for space group  $R\bar{3}c$ and lattice parameters $a$ = 7.8915(8) {\AA} and $c$ = 20.788(6) {\AA}. Violet curves are the difference between data and calculated patterns. Inset: crystal picture. (b) Zero-field resistivity with current parallel to the [100] axis. (c) Inverse magnetic susceptibility $H/M$ vs. $T$ for magnetic field $H\|$[001] (blue), [100] (red), and a polycrystalline average (violet). Solid lines are high temperature Curie-Weiss fits, with Weiss temperatures of $-360$ K, $-20$ K and $-100$ K for $H\|$[001], [100], and the polycrystalline average, respectively (see text). Left inset: the low-temperature $H/M$ vs. $T$ with a linear fit for the polycrystalline average. Right inset: the low-temperature $M/H$ vs. $T$ for $H\|$[100].}
	\end{figure}
Here we report the discovery of CeIr$_3$Ge$_7$, a new intermetallic compound without Kondo effect and no geometric frustration, with a remarkably low AFM ordering temperature $T_{\rm N}$ = 0.63 K. This is one of several $R$ = Ce or Yb compounds we recently discovered in the $R$$T$$_3$$M$$_7$ (1-3-7) class of compounds with $T$ = transition metal and $M$ = group 14 element\cite{BinodPRX2017, Binod2017, BinodYIG2017}, a family of rhombohedral intermetallics with the ScRh$_3$Si$_7$ structure type\cite{Lorenz,Chabot}. The $R$ sublattice forms a cubic structure, with nearest-neighbor $d_{R-R}$ around 5.7 {\AA}. Strong electron correlations in the Yb members of this family result in HF behavior and ferromagnetic or AFM ordering at temperatures as high as 7.5 K. Remarkably, CeIr$_3$Ge$_7$ is weakly correlated, and even with similar $d_{R-R}$, the ordering temperature is much smaller than in the HF Yb analogues. This seems uncommon when compared to other HF systems where the Yb compounds usually order at lower temperatures than the Ce counterparts, due to deeper localization of the 4$f$ electron and the larger strength of the spin-orbit coupling in the former \cite{Flouquet}. Such a comparison could be tenuous since we neglect all the details of band structures. No frustration is present in CeIr$_3$Ge$_7$, as the Weiss temperature in the limit of absolute zero is close to $T_{\rm N}$. This system is a good metal, with residual resistivity values $\rho_0 \sim$ 20 $\mu \Omega$ cm and a residual resistivity ratio RRR = $\rho(300K)/\rho_0 \sim$ 5. No Kondo correlations are apparent as most of the magnetic entropy is released below $T_{\rm N}$. In the absence of Kondo effect or frustration, the low $T_{\rm N}$ in CeIr$_3$Ge$_7$ is attributed to the large distance $d_{R-R}$. This points to the potential of the 1-3-7 family  to reveal Ce or Yb compounds with low ordering temperatures, which, in turn, may be easily tuned towards absolute zero transitions and quantum critical regimes. 
	
CeIr$_{3}$Ge$_{7}$ crystallizes in the $R\bar{3}c$ rhombohedral ScRh$_{3}$Si$_{7}$ structure type\cite{Lorenz,Chabot}. The very few 1-3-7 compounds known so far are $R$Au$_3$Ga$_7$ ($R$ = Gd-Yb)\cite{Cordier,VERBOVYTSKYY}, non-magnetic $R$Au$_3$Al$_7$ ($R$ = Ce-Sm, Gd-Lu) \cite{Latturner}, and magnetic Eu(Rh,Ir)$_3$Ge$_7$ \cite{Falmbigl}. Recently we discovered the first magnetic Ce and Yb 1-3-7 compounds. All of the newly discovered compounds were synthesized in single crystal form using a self-flux growth method\cite{Remeika, BinodCoM}, with details described elsewhere\cite{BinodPRX2017}. Single crystal x-ray diffraction measurements confirm the ScRh$_3$Si$_7$ structure type and verify the purity and stoichiometry of these compounds. The details of the x-ray diffraction and experimental methods are described in the Supplementary Material \cite{Supplement}. A nonmagnetic analogue YIr$_{3}$Ge$_{7}$ polycrystalline sample was prepared by arc melting. Figure \ref{Fig1}(a) shows a powder x-ray pattern and structural refinement for CeIr$_3$Ge$_7$ with a photo of a crystal shown in the inset. In this rhombohedral crystal structure, the $R$ atoms form a cubic sublattice\cite{BinodPRX2017}, with the body diagonal of the cube parallel to the $c$ axis of the equivalent hexagonal unit cell. Notably, the distances $d_{R-R} \sim$ 5.7 {\AA} are larger than in many magnetic $R$ intermetallics, but do not change significantly for $R$ = Ce or Yb in the 1-3-7 structure. This observation becomes most relevant when trying to explain the low ordering temperature $T_{\rm N}$ = 0.63 K in CeIr$_3$Ge$_7$. Several scenarios may in principle result in low $T_{\rm N}$ in Ce compounds, such as the Kondo effect, frustration or exchange coupling competition, weak exchange due to large $d_{\rm {Ce}-\rm {Ce}}$. The following discussion is based on evidence against most, if not all, of these scenarios in CeIr$_3$Ge$_7$, rendering this compound a unique non-Kondo metal with extremely low ordering temperature.

The $\mu_0 H$ = 0 resistivity measurements (Fig. \ref{Fig1}~(b)) show that CeIr$_3$Ge$_7$ is a good metal, with a RRR = 5 and residual resistivity $\rho_0 \sim$ 20 $\mu \Omega$cm. However, upon cooling from room temperature, the resistivity is linear in temperature, and no signatures of Kondo correlations are apparent. The lack of Kondo effect will be further corroborated by the specific heat data shown later. For now, we turn to the magnetic susceptibility measured along ($H\|$[001]) and perpendicular ($H\|$[100]) to the $c$ axis of the equivalent hexagonal unit cell. The inverse susceptibility $H/M$ (Fig. \ref{Fig1}(c)), measured up to 600 K, reveals large easy-plane CEF anisotropy. The average susceptibility is calculated as $M_{ave} = (M_{001} + 2 M_{100})/3$. Fits to the Curie-Weiss law at high temperatures are shown in solid black lines. The experimental effective moment $\mu^{exp}_{eff}$ extracted from the fit of the average susceptibility (violet, Fig. \ref{Fig1}(c)) is $\mu^{exp}_{eff}$ = 2.52 $\mu_{\rm B}$/Ce$^{3+}$, pointing to localized, trivalent Ce ions in CeIr$_3$Ge$_7$, since the expected Ce$^{3+}$ calculated effective moment $\mu^{calc}_{eff}$ = 2.52 $\mu_{\rm B}$/Ce$^{3+}$ is virtually identical to the experimental value. The negative Weiss temperatures indicate AFM correlations. The $H/M$ data deviate from the Curie-Weiss law due to CEF splitting of the $J = 5/2$ multiplet. The deviation indicates a separation of the first excited CEF doublet of $\sim 400$ K, consistent with the CEF calculations which will be reported elsewhere \cite{Jacintha}. 

\begin{figure}[hb!]
	\includegraphics[width=1\columnwidth]{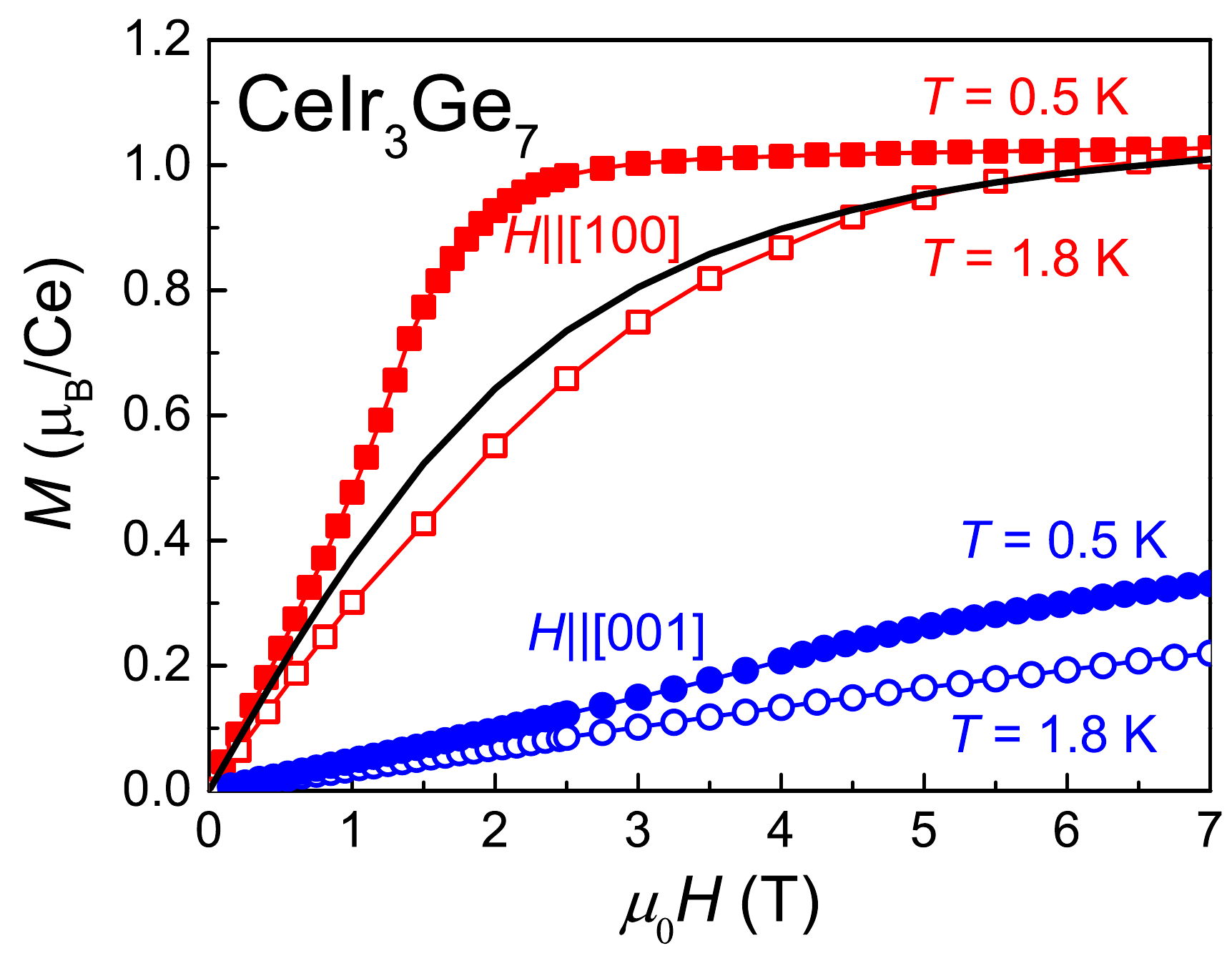}
	\caption{\label{Fig2} CeIr$_3$Ge$_7$ $M$~vs.~$H$ isotherms for $T =$ 0.5 K (full symbols) and 1.8 K (open symbols) for $H\|[001]$ (blue circles) and $H\|[100]$ (red squares). The solid black line is the Brillouin function for $T$ = 1.8 K, with a scaled magnitude.}
\end{figure}

In the $T \rightarrow 0$ limit, the inverse susceptibility intercept with the temperature axis is around $-2$ K for $M_{ave}$ (left inset, Fig. \ref{Fig1}(c)), comparable to the low ordering temperature $T_{\rm N} \sim$ 0.6 K indicated by the cusp in $M/H$ vs. $T$ for $H\|$[001] (right inset, Fig. \ref{Fig1}(c)). These observations can be reconciled by considering the Weiss temperatures at $T \rightarrow 0$ to reflect the exchange coupling $J$$_{ex}$, which consequently indicates that $J$$_{ex}$ is inherently small in CeIr$_3$Ge$_7$. At high $T$, the Weiss temperatures are enhanced by the large CEF effects. We show the $M(H)$ isotherms in Fig. \ref{Fig2} for $H\|$[100] (red symbols) and $H\|$[001] (blue symbols), in the ordered state $T$ $=$ 0.5 K (full symbols) and the paramagnetic state $T$ = 1.8 K (open symbols). A Brillouin function at 1.8 K along [100] direction, shown as a solid black line in Fig. \ref{Fig2}, agrees very well with the experimental data, indicating that the measured magnetization mostly comes from isolated paramagnetic moments. The $M(H)$ measurements confirm the in-plane and out-of-plane magnetic anisotropies, and, more quantitatively, are in good agreement with the calculated moments of 0.91 $\mu_{B}$/Ce and 0.33 $\mu_{B}$/Ce along the easy ([100]) and hard ([001]) directions, respectively \cite{Jacintha}. A magnetic field close to $\mu_0 H$ = 1.7 T is required for saturation in the easy direction (squares, Fig. \ref{Fig2}), while a linear extrapolation of $M(H\parallel[001])$ suggests a magnetic field in excess of 20 T is needed to reach saturation in the hard direction. 

\begin{figure}
	\includegraphics[width=1\columnwidth]{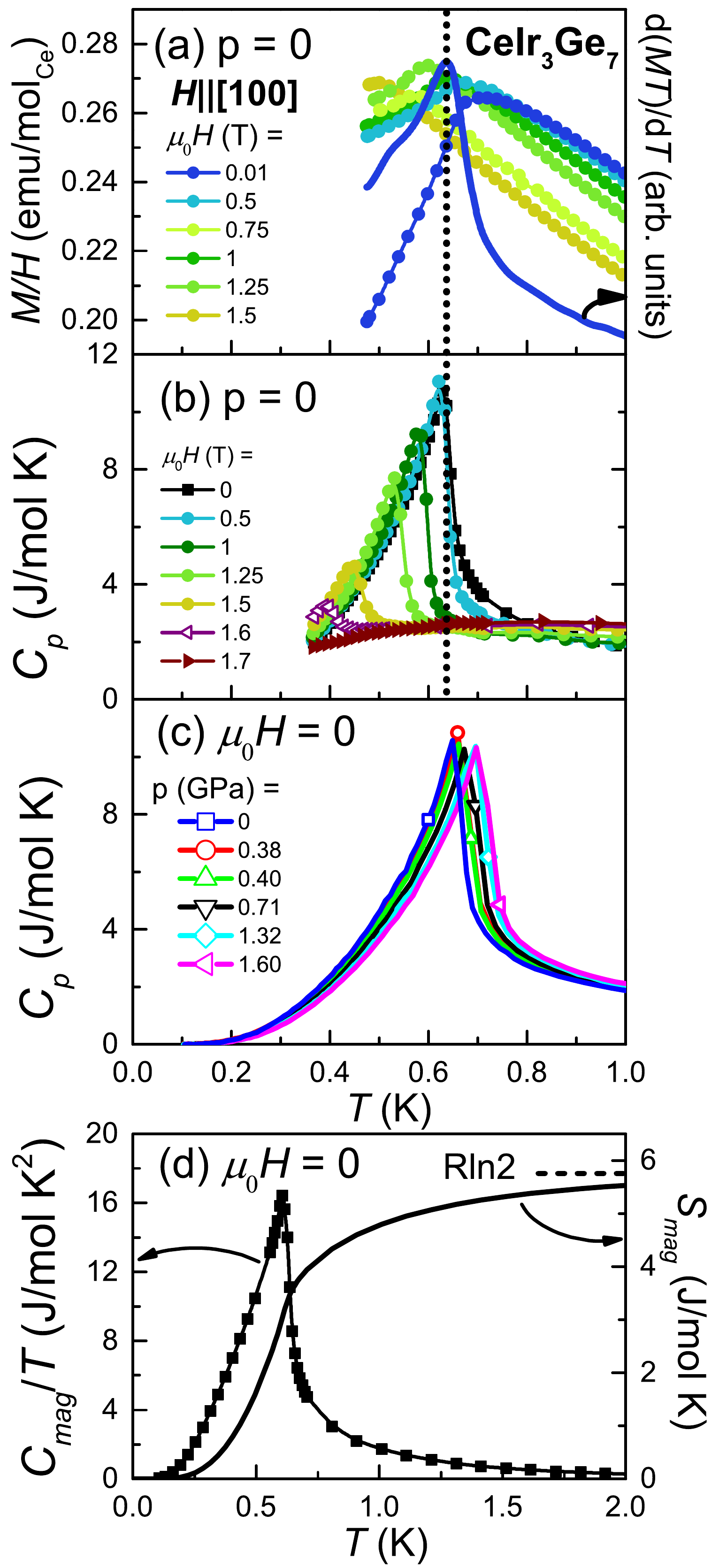}
	\caption{\label{Fig3} (a) Left axis: magnetic susceptibility $M/H$ \textit{vs.} $T$. Right axis: d$(MT)$/d$T vs. T$ (solid line) for $\mu_{0} H$ $=$ 0.01 T. (b) Left axis: specific heat $C_p$ \textit{vs.} $T$ (symbols) for different magnetic fields. The vertical dashed line through (a) to (b) marks $T_{\rm N}$ at zero field. (c) Temperature dependent specific heat for different pressures. (d) Left axis: magnetic contribution to the specific heat $C_{mag}/T$ \textit{vs.} $T$. Right axis: magnetic entropy $S_{mag}$ \textit{vs.} $T$. $C_{mag}/T$ = $C_p/T$(CeIr$_3$Ge$_7$) - $C_p/T$(YIr$_3$Ge$_7$), where YIr$_3$Ge$_7$ is a nonmagnetic analogue.}
\end{figure}

Because of this extremely large anisotropy, and the large magnetic field scale in the hard ([001]) direction, we focus next only on the field dependence of the ordering temperature $T_{\rm N}$ for the easy direction $H\|$[100], as illustrated by the $M/H$ and specific heat $C_p$ data in Fig. \ref{Fig3}. For AFM systems, a peak in $C_p$ at $T_{\rm N}$ is expected to correspond to a peak in d$(MT)$/d$T$\cite{Fisher}, and this is illustrated for $\mu_0 H = $ 0.01 T (solid line, right axis) in Fig. \ref{Fig3}(a). Both ambient pressure $M/H$ and $C_p$ measurements (\ref{Fig3}(a-b)) reveal the expected suppression of $T_{\rm N}$ with increasing $H$, such that above $\mu_0 H = $ 1.7 T, no peak can be resolved above 350 mK. Consistent with the zero field resistivity data in Fig. \ref{Fig1}(b), the specific heat data show an entropy release of $\sim$ 2$/$3 R$\ln$2 at $T_{\rm N}$ (solid line, right axis in Fig. \ref{Fig3}c), reaffirming the absence of both Kondo effect and strong correlations in CeIr$_3$Ge$_7$. 
To further rule out the presence of Kondo screening, one can consider the analysis by de Jongh and Miedema \cite{Jongh}, which shows that an entropy release of 15$\%$-40$\%$ of R$\ln$2 above $T_{\rm N}$ can be expected in AFM systems around $T_{\rm N}$. Indeed this is reflected in the $H =$ 0 magnetic entropy plot of CeIr$_3$Ge$_7$ in Fig. \ref{Fig3}(b) (black line). Furthermore, the same model indicates that the $C_p$ contribution from Kondo is much weaker than that from classical intersite fluctuations. Within a Heisenberg model, one expects that, far above $T_{\rm N}$, the leading term in $C_p (T)$ is proportional to 1$/T^2$, \textit{i.e.}, $C_p/T$ $\sim$ $1/T^3$. Upon applying a magnetic field (Fig. \ref{Fig3}(b)), the dispersion of the magnons changes. In an AFM system the energy gap at Q $=$ 0 decreases and disappears at $\mu_0 H = \mu_0 H_c \approx$ 1.7 T. This results in a large increase of the low energy magnon-like excitations, which, in turn, shows up as a strong increase of $C_p$ near and above $T_{\rm N}$. For $H > H_c$ (full right triangles, Fig. \ref{Fig3}(b)), a gap reopens in the magnon excitation spectra, and the specific heat evolves towards a broad anomaly related to the dominant Zeeman splitting. 

	\begin{figure}
			\includegraphics[width=1\columnwidth]{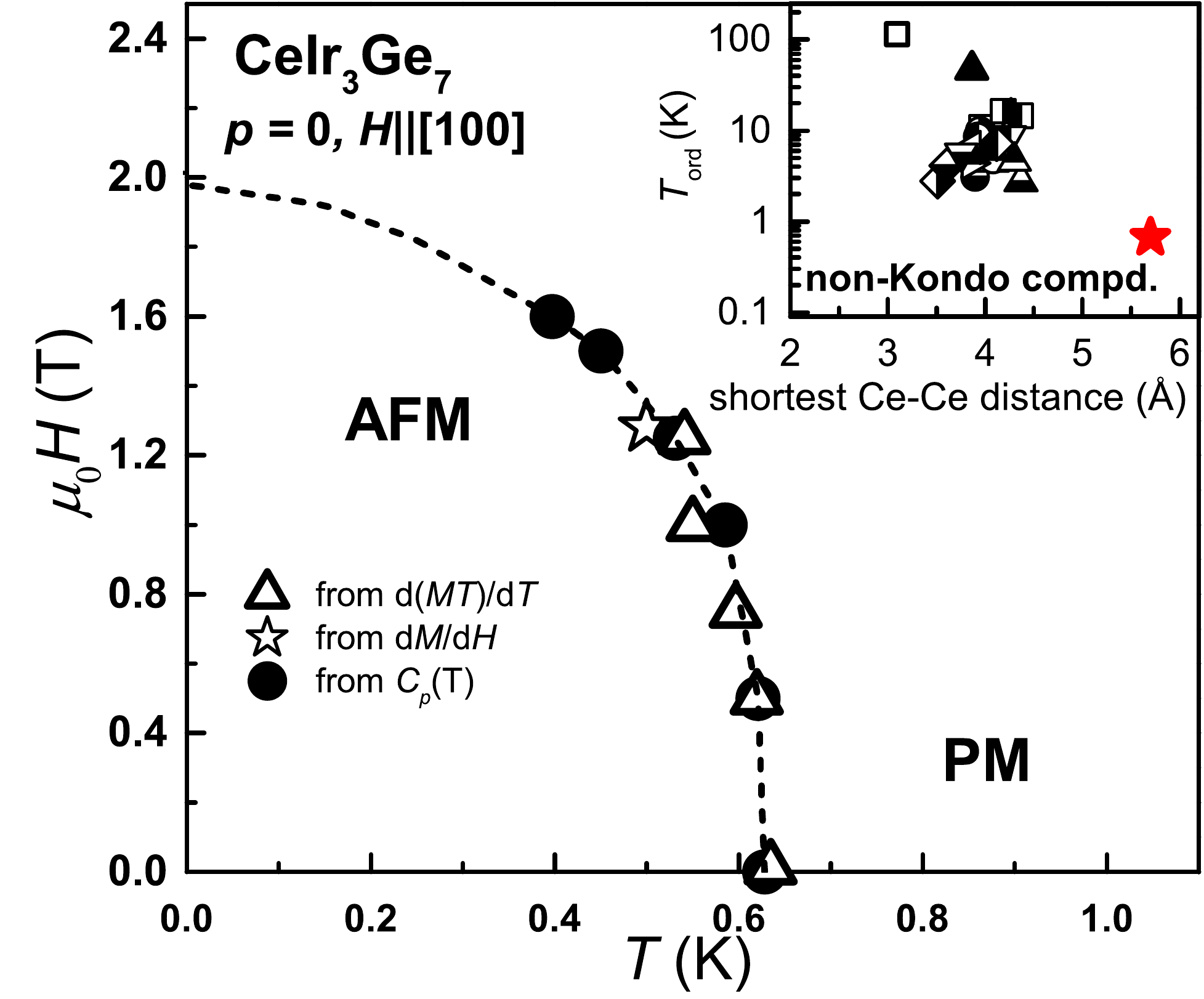}
			\caption{\label{Fig4} $T - H$ phase diagram of CeIr$_3$Ge$_7$ at ambient pressure with $H\|$[100]. Inset: ordering temperature $T_{\rm {ord}}$ vs. the shortest Ce-Ce distance in known intermetallic non-Kondo Ce compounds. Different symbols refer to different compounds tabulated in Supplemental Material \cite{Nicklas,von Blanckenhagen,Kraft,Mhlungu,Ajeesh,Huang,Lenkewitz,Grytsiv,Gignoux,Oner,Huang2,Klicpera,Loidl,Luo,Nakotte,Schille,Shigetoh,Baumbach,Shaheen,Thamizhavel,Sereni}.}
	\end{figure}

Complementary to the field dependence, the pressure dependence of the specific heat (Fig. \ref{Fig3}(c)) underlines the conclusion of small or negligible Kondo correlations: $T_{\rm N}$ increases linearly with pressures up to 1.6 GPa at a rate of $dT_{\rm N}/dp = 3.71\times10^{-2}$ K/GPa. The increase of $T_{\rm N}$ under pressure, if only being ascribed to a volume effect, can be understood in the framework of the Doniach phase diagram \cite{Doniach}. The positive, yet very small, slope of $T_{\rm N}(p)$ for CeIr$_3$Ge$_7$ suggests that this compound is located at far left of $T_{\rm N}^{max}(J_{ex})$ in the Doniach diagram \cite{Doniach}. The $T-H$ phase diagram in Fig. \ref{Fig4} summarizes the $T_{\rm N}$ dependence on field at ambient pressure.


Among non-Kondo magnetic Ce compounds, CeIr$_3$Ge$_7$ stands out (red in inset of Fig. \ref{Fig4}) together with CeRh$_3$B$_2$ (open square in inset of Fig. \ref{Fig4}). The latter orders ferromagnetically with a remarkably large Curie temperature ($\sim$ 115 K) due to the enhancement of the exchange interaction from the $J$ = 7/2 multiplet, despite short $d_{\rm {Ce}-\rm {Ce}} =$ 3.096~{\AA} \cite{Givord2007}.  Of note is the compound Ce$_4$Pt$_{12}$Sn$_{25}$ (Ref.~\onlinecite{Kurita}), which appears to have a record low $T_{\rm N} =$ 0.18 K and represents a Kondo lattice in the small exchange limit of the Doniach phase diagram. In this case, however, the extremely low $T_{\rm N}$ is a result of the large $d_{\rm {Ce}-\rm {Ce}} \sim$ 6.14 {\AA}, weak Kondo screening just above $T_{\rm N}$ (marked by a tail in the magnetic specific heat peak just above the ordering), and weak geometric frustration due to the three-fold point symmetry of the Ce site \cite{White}. Except for the large $d_{\rm {Ce}-\rm {Ce}} \sim$ 5.7 {\AA}, none of these effects are at play in CeIr$_3$Ge$_7$: the low temperature Weiss temperatures (Fig. \ref{Fig1}(c) inset) are comparable with $T_{\rm N}$, ruling out significant frustration effects; the specific heat peak (Fig. \ref{Fig3}(b)) terminates abruptly at $T_{\rm N}$, and $\rho(T)$ decreases linearly with temperature before it levels off at $\rho_0$ at the lowest temperatures (Fig. \ref{Fig1}(b)), therefore no Kondo screening signatures are apparent. Ce and Yb magnetic (trivalent) compounds are often thought as electron-hole analogues. In metals, the exchange coupling decreases from Ce to Yb, and therefore larger ordering temperatures are expected in the former compared to the latter. So, in the absence of Kondo effect or frustration, CeIr$_3$Ge$_7$ should have a magnetic ordering temperature larger than its Yb analogue. What we find is that YbIr$_3$Ge$_7$ is in fact an HF ferromagnet \cite{BinodYIG2017}, and this should further suppress the magnetic order to temperatures well below $T_{\rm N}$(Ce). Instead, the Curie temperature for this HF ferromagnet is large $T_{\rm C} \sim$ 2.4 K, despite the nearly identical $d_{R-R}$ in both the Ce and Yb analogues. This may reflect that the details of the band structure near the Fermi surface of Ce and Yb analogues plays an important role of magnetism. In summary, CeIr$_3$Ge$_7$ in particular and more generally the 1-3-7 family of magnetic compounds provide a fertile ground for exploring magnetic correlations and the competition among various energy scales (RKKY, Kondo, CEF) which could result in novel quantum critical regimes. In addition, their rhombohedral structure allows for very weak coupling between the Ce atoms in a good metallic environment, similar to what was observed in YbPt$_2$Sn \cite{Jang2015}. This is an excellent precondition for metallic magnets that can be used for adiabatic demagnetization cooling below 2~K, instead of insulating paramagnetic salts.

\textbf{Acknowledgements}
We thank J. Simone, J. D. Thompson, C. Geibel, and W. Hu for fruitful discussions. Work at Rice University was supported by the Gordon and Betty Moore Foundation EPiQS Initiative through grant GBMF 4417. The work at University of Texas at Dallas was supported by NSF-DMR 1700030. This research is funded in part by a QuanEmX grant from ICAM and the Gordon and Betty Moore Foundation through Grant GBMF5305 to Binod Rai. EM acknowledges travel support to Max Planck Institute in Dresden, Germany from the Alexander von Humboldt Foundation Fellowship for Experienced Researchers.

\end{document}